\documentclass[aps,amssymb,pra,showpacs,twocolumn]{revtex4}
\usepackage{epsf}
\usepackage{amsmath}
\usepackage{graphicx}
                             
\begin{document}                                                       
                                            
\title{Underbarrier interference and Euclidean resonance}

\author{B. Ivlev} 

\affiliation
{Instituto de F\'{\i}sica, Universidad Aut\'onoma de San Luis Potos\'{\i}, San Luis Potos\'{\i}, Mexico,\\
Department of Physics and Astronomy and NanoCenter, University of South Carolina, Columbia, South Carolina, USA}


\begin{abstract}
 
Quantum tunneling from a thin wire or a thin film through a static potential barrier in a zero magnetic field is studied. 
The wire or the film should satisfy a condition of transverse quantization of levels and be inhomogeneous. Depending on a 
form of the inhomogeneity the tunneling scenario can dramatically differ from the conventional scheme of an exponential 
decay of a probability density inside the barrier. This happens due to interference of various undrebarrier paths, which 
are collected at some points giving rise to a trans-barrier state of a large amplitude. As a result, the tunneling 
probability through an almost classical barrier can be not exponentially small. This is a phenomenon of Euclidean resonance 
studied earlier for tunneling across nonstationary barriers.

\end{abstract} \vskip 1.0cm
   
 
\maketitle
\section{Introduction}
Quantum tunneling through a one-dimensional static potential barrier is described by the theory of Wentzel, Kramers, and
Brillouin (WKB) \cite{LANDAU} if the barrier is not very transparent. For a two-dimensional static barrier the
semiclassical approach $\psi\sim\exp[iS(x,y)/\hbar]$ for the wave function is appropriate, where $S(x,y)$ is the 
classical action.         

The main contribution to the tunneling probability comes from the extreme path in $\{x,y\}$ plane linking two classically 
allowed regions. The path can be parameterized  as a classical trajectory in imaginary time. See, for example, 
Refs.~\cite{LIFSHITZ,POKR,VOL,STONE,COLEMAN1,COLEMAN2,MILLER,SCHMID1,SCHMID2}. The trajectory is a solution of Newton's 
equation in imaginary time and it is locally orthogonal to borders of the linked classical regions. The classical action 
becomes imaginary under the barrier and it is well defined at each point of the trajectory. 

According to that scenario, a phase of the wave function is zero under the barrier and one can apply WKB formalism along
the path. But sometimes a phase of the underbarrier wave function may be not zero. This can happen, for example, in 
tunneling from a state of a quantum wire, associated with a finite velocity along the wire.

The underbarrier phase can lead to unexpected peculiarities of the tunneling process due to an interference under the barrier.
For instance, various paths are collected at certain points resulting in singularities of the action $S(x,y)$. In 
principle, appearance of the singularity is not surprising since it may be related to an exit point, which exists even 
without an underbarrier phase. At such a point the semiclassical approximation breaks down and branching occurs. 

\begin{figure}
\includegraphics[width=6.0cm]{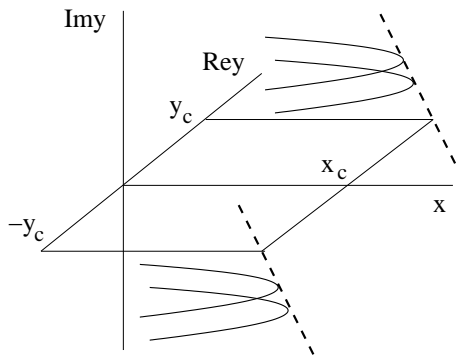}
\caption{\label{fig1} The caustics are marked by the dashed curves. They pin the physical $\{x,y\}$ plane at the points
$\{x_{c},\pm y_{c}\}$. The classical trajectories relate to complex variables and are reflected from the caustics.} 
\end{figure}

However, with a phase under the barrier the extreme path no more lies in the physical $\{x,y\}$ plane but is shifted towards
complex values. Complex classical trajectories are reflected from the certain curves, which are caustics \cite{LANDAU1} (see
also Refs.~\cite{SCHMID1,SCHMID2,DYK}). Such a case is shown in Fig.~\ref{fig1}, where the caustics pin the physical 
$\{x,y\}$ plane at the points $\{x_{c},\pm y_{c}\}$. This situation may correspond to tunneling from a inhomogeneous quantum 
wire placed along the line $x=0$. In that case an electron has a velocity at least at some parts of the wire, which results 
in a phase under the barrier. We emphasize that the barrier is static and a magnetic field is zero. 

At the points $\{x_{c},\pm y_{c}\}$ of singularities of the action $S(x,y)$ a phenomenon of branching occurs. The original 
branch, which starts at the line $x=0$, is matched with the certain trans-barrier branch at the singularity points. The 
trans-barrier state has the same current direction, along $y$, as the state at the wire. Under some conditions, the amplitude 
of the trans-barrier state can be not exponentially small giving rise to a possibility of penetration through classical 
potential barriers. This is a phenomenon of Euclidean resonance studied before in tunneling through nonstationary barriers 
\cite{IVLEV2,IVLEV3,IVLEV4,IVLEV5}, where the underbarrier phase was created by quanta emission. In our case the 
underbarrier phase is driven by the current in the quantum wire. Various underbarrier paths interfere resulting in the 
singularity points. A formation of a phase under the barrier is a necessary condition for the phenomenon. An idea of 
Euclidean resonance in tunneling through a static barrier was proposed in Ref.~\cite{IVLEV6}.

Below we consider the phenomenon of Euclidean resonance for tunneling from a inhomogeneous quantum wire, in a zero magnetic 
field, through a potential barrier created by an applied static electric field. 
\section{FORMULATION OF THE PROBLEM}
\label{formul}
We consider an electron localized in the $\{x,y\}$ plane and described by the static Schr\"{o}dinger equation 
\begin{eqnarray} 
\nonumber
&&-\frac{\hbar^{2}}{2m}\left(\frac{\partial^{2}\psi}{\partial x^{2}}+\frac{\partial^{2}\psi}{\partial y^{2}}\right)
-\hbar\sqrt{\frac{2u_{0}}{m}}\sqrt{1+\alpha^{2}(y)}\hspace{0.1cm}\delta(x)\psi\\
\label{1}
&&-{\cal E}_{0}|x|\psi=E\psi.
\end{eqnarray}
The positive function $\alpha(y)$ is even and $\alpha(\infty)\rightarrow 0$. The $\delta$ well in Eq.~(\ref{1})
describes the long quantum wire placed at the position $x=0$ and with the discrete energy level 
$-u_{0}[1+\alpha^{2}(y)]$. In other words, the wire is inhomogeneous along its length attracting the electron to 
the domain of a finite $y$. Tunneling occurs through the triangular potential barrier created by the static electric field 
${\cal E}_{0}$. For simplicity we use the even potential $-{\cal E}_{0}|x|$ to get the problem symmetric with respect to $x$. 

Below we measure $x$ and $y$ in the units 
of $u_{0}/{\cal E}_{0}$. The large semiclassical parameter is
\begin{equation} 
\label{2}
B=\frac{u_{0}\sqrt{2mu_{0}}}{\hbar{\cal E}_{0}}.
\end{equation}
In the new units the Schr\"{o}dinger equation reads
\begin{equation} 
\label{3}
-\frac{1}{B^{2}}\left(\frac{\partial^{2}\psi}{\partial x^{2}}+\frac{\partial^{2}\psi}{\partial x^{2}}\right)
-\frac{2}{B}\sqrt{1+\alpha^{2}(y)}\hspace{0.1cm}\delta(x)\psi-|x|\psi=\lambda\psi,
\end{equation}
where $E=u_{0}\lambda$. In the limit of a large $B$ the border of the continuous spectrum on $y$ in the isolated quantum 
wire is $\lambda=-1$. The states in the wire with $\lambda<-1$ are discrete. Below we consider $\lambda=-1$.
\section{HAMILTON-JACOBI EQUATION}
\label{ham}
In the semiclassical approximation a wave function can be written in the form \cite{LANDAU} 
\begin{equation} 
\label{4}
\psi(x,y)\sim\exp\left[iB\sigma(x,y)\right],
\end{equation}
where $\hbar B\sigma$ is the classical action. $\sigma$ satisfies the Hamilton-Jacobi equation at $x\neq 0$
\begin{equation} 
\label{5}
\left(\frac{\partial\sigma}{\partial x}\right)^{2}+\left(\frac{\partial\sigma}{\partial y}\right)^{2}-|x|=-1.
\end{equation}
The equation (\ref{5}) should be supplemented by a boundary condition at $x=0$. The function $\sigma(x,y)$ is continuous
at the line $x=0$. Then, as follows from Eq.~(\ref{5}), $(\partial\sigma/\partial x)^{2}$ has the same values at $x=\pm 0$. 
In the approximation (\ref{4}), as follows from Eq.~(\ref{3}), this corresponds to the boundary condition at $x=+0$
\begin{equation} 
\label{6}
\frac{\partial\sigma(x,y)}{\partial x}\bigg|_{x=0}=i\sqrt{1+\alpha^{2}(y)}.
\end{equation}
A general integral of the Hamilton-Jacobi equation (\ref{5}) can be obtained by the method of variation of constants
\cite{LANDAU2}. At a positive $x$ the solution of Eq.~(\ref{5}), satisfying the condition (\ref{6}), has the form
\begin{eqnarray} 
\nonumber
&&\sigma(x,y)=y\alpha[u(x,y)]+i\int^{x}_{0}dx_{1}\sqrt{\alpha^{2}[u(x,y)]+1-x_{1}}\\
\label{7}
&&-\int^{u(x,y)}_{0}du_{1}u_{1}\frac{\partial\alpha(u_{1})}{\partial u_{1}},
\end{eqnarray}
where the function $u(x,y)$ obeys the equation
\begin{equation} 
\label{8}
u(x,y)-y=i\int^{x}_{0}dx_{1}\frac{\alpha[u(x,y)]}{\sqrt{\alpha^{2}[u(x,y)]+1-x_{1}}}.
\end{equation}
Eq.~(\ref{8}) is the condition $\partial\sigma/\partial u=0$, which is equivalent to independence of $\sigma$ on
``constant'' $u(x,y)$. In this formalism the relations hold
\begin{eqnarray}
&&\label{9}
\frac{\partial\sigma(x,y)}{\partial x}=i\sqrt{\alpha^{2}[u(x,y)]+1-x},\\
\label{10}
&&\frac{\partial\sigma(x,y)}{\partial y}=\alpha[u(x,y)].
\end{eqnarray}
As follows from Eq.~(\ref{8}), $u(0,y)=y$ and therefore $\partial\sigma(0,y)/\partial y=\alpha(y)$.
\section{SOLUTIONS}
\label{sol}
\begin{figure}
\includegraphics[width=6cm]{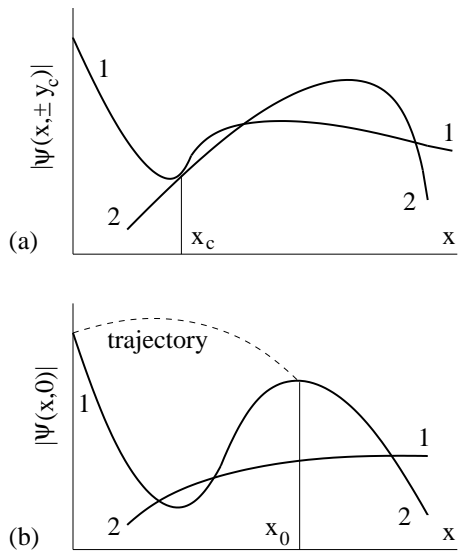}
\caption{\label{fig2} (a) The conventional (1-1) and the trans-barrier (2-2) branches touch each other at the caustic 
points $\{x_{c},\pm y_{c}\}$, where the reconnection occurs. (b) After the reconnection there is no violation of the 
semiclassical conditions along the hybridized branch 1-2. The dashed curve (trajectory) indicates the bypass through
the complex plane.}
\end{figure}

An analysis of Eqs.~(\ref{7}-\ref{10}) shows that at a not small $|y|$ there is a branch of the action $\sigma$ generic with 
the conventional WKB branch under the barrier. It starts at the enter into the barrier ($x=0$) and is indicated in 
Fig.~\ref{fig2}(a) as 1-1. 

There is also another branch indicated as 2-2 in Fig.~\ref{fig2}(a). The top point of this branch follows the classical 
trajectory
\begin{equation} 
\label{11}
x=x_{0}+\frac{y^{2}}{4(x_{0}-1)}.
\end{equation}
The parameter $x_{0}$ is calculated below. Within the semiclassical approximation used the modulus of the wave function is 
a constant on the trajectory (\ref{11}) and the branch 2-2 is more smeared out at a larger $|y|$. At the certain 
$y=\pm y_{c}$ the branch 2-2 touches the branch 1-1 at the point $x=x_{c}$, as demonstrated in Fig.~\ref{fig2}(a). At that 
point the semiclassical approach breaks down and the branch reconnection occurs. The result of the reconnection is shown in 
Fig.~\ref{fig2}(b) at $y=0$. Along the hybridized branch 1-2 in Fig.~\ref{fig2}(b) the semiclassical approximation is not 
broken. 

The tunneling probability is
\begin{equation} 
\label{12}
w=\frac{|\psi(x_{0},0)|^{2}}{|\psi(0,0)|^{2}}\sim\exp\left[-2{\rm Im}\sigma({x_{0},0})\right].
\end{equation}
The function $\sigma(x,0)$ is determined from the equations
\begin{eqnarray} 
\nonumber
&&\sigma(x,0)=i\int^{x}_{0}dx_{1}\sqrt{\alpha^{2}[iv(x)]+1-x_{1}}\\
\label{13}
&&-i\int^{v(x)}_{0}dv_{1}v_{1}\frac{\partial\alpha(iv_{1})}{\partial v_{1}},
\end{eqnarray}
where the function $v(x)=-iu(x,0)$ is defined as
\begin{equation} 
\label{14}
v(x)=\int^{x}_{0}dx_{1}\frac{\alpha[iv(x)]}{\sqrt{\alpha^{2}[iv(x)]+1-x_{1}}}.
\end{equation}
We specify the form
\begin{equation} 
\label{15}
\alpha(y)=\alpha_{0}\exp\left(-\frac{y^{2}}{a^{2}}\right),\hspace{1cm}0<\alpha_{0}.
\end{equation}
The parameter $\alpha_{0}=2$ is taken below. It is not difficult to do calculations on the basis of Eqs.~(\ref{13}) and 
(\ref{14}). The tunneling probability (\ref{12}) becomes of the form
\begin{equation} 
\label{16}
w\sim\exp\left[-2.1(a_{R}-a)B\right],
\end{equation}
where the resonant value is $a_{R}\simeq 39.5$. The condition $a=a_{R}$ corresponds to Euclidean resonance, when the
tunneling probability is not exponentially small. Eq.~(\ref{16}) is valid if $(a_{R}-a)$ is not large. The top point in 
Fig.~\ref{fig2}(b) is determined from the condition of zero of the expression $\alpha^{2}[iv(x)]+1-x$, which is quadratic 
close to zero. The calculations lead to $x_{0}\simeq 16$. At $a=a_{R}$, as follows from Eqs.(\ref{7}) and (\ref{8}), 
$x_{c}\simeq 5.6$ and $y_{c}\simeq 0.14$. 

\section{CLASSICAL TRAJECTORIES IN IMAGINARY TIME}
\label{traj}
The tunneling probability within the exponential approach (\ref{12}) can be obtained by the method of classical complex 
trajectories in imaginary time $t=i\tau$. The coordinate $x(\tau)$ remains real in imaginary time but the other coordinate 
becomes imaginary $y(\tau)=-i\eta(\tau)$. This type of complex coordinates was used in magnetotunneling \cite{BLATT}. See
also Refs.~\cite{DYK,GOROKH}. We consider tunneling from the $\delta$ well ($\tau=\tau_{0}$) to the top point in 
Fig.~\ref{fig2}(b) ($\tau=0$). One can say that the trajectory provides a bypass through the complex plane.

We choose the trajectory to get physical values $y(0)=0$ and $x(i\tau_{0})=0$. The probability of tunneling 
\begin{equation} 
\label{17}
w\sim\exp\left(-A\right)
\end{equation}
depends on the parameter
\begin{equation} 
\label{18}
A=2B\hspace{0.1cm}{\rm Im}\left[\sigma(x_{0},0)-\sigma(0,0)\right].
\end{equation}
We use the notations $x_{0}=x(0)$ and $\eta_{0}=\eta(i\tau_{0})$. The trajectory method allows to calculate the part $A_{0}$ 
of the total action (\ref{18}) only. This part
\begin{equation} 
\label{19}
{A}_{0}=2B\hspace{0.1cm}{\rm Im}\left[\sigma(x_{0},0)-\sigma(0,-i\eta_{0})\right].
\end{equation}
is expressed through the trajectory $\{x(\tau),\eta(\tau)\}$ 
\begin{equation} 
\label{20}
A_{0}=2B\int^{\tau_{0}}_{0}\left[\frac{1}{4}\left(\frac{\partial x}{\partial\tau}\right)^{2}-
\frac{1}{4}\left(\frac{\partial\eta}{\partial\tau}\right)^{2}-x+1\right]d\tau
\end{equation}
The coordinates $x(\tau)$ and $\eta(\tau)$ in Eq.~(\ref{20}) satisfy the classical equations of motion 
\begin{equation} 
\label{21}
\frac{1}{2}\frac{\partial^{2}x}{\partial\tau^{2}}=-1,\hspace{0.5cm}
\frac{1}{2}\frac{\partial^{2}\eta}{\partial\tau^{2}}=0.
\end{equation}
The solutions are
\begin{equation} 
\label{22}
x(\tau)=1+\alpha^{2}(-i\eta_{0})-\tau^{2},\hspace{0.3cm}\eta(\tau)=-2\alpha(-i\eta_{0})\tau.
\end{equation}
The trajectory in the form (\ref{11}) follows from Eqs.~(\ref{22}).

The trajectory terminates at the unphysical (complex) point $x=0$, $y=-i\eta_{0}$. One should connect this point with a 
physical one, say, $x=0$, $y=0$. So the total action is $A=A_{0}+A_{1}$, where
\begin{equation} 
\label{23}
A_{1}=2B{\rm Im}\left[\sigma(0,-i\eta_{0})-\sigma(0,0)\right].
\end{equation}
One can find the action (\ref{23}) by a direct solution of the Hamilton-Jacobi equation (\ref{5})
if to put $x=0$ and to use the condition (\ref{6}). The result is
\begin{equation} 
\label{24}
A_{1}=-2B\int^{\eta_{0}}_{0}d\eta\alpha(-i\eta).
\end{equation}
Finally, the tunneling probability is 
\begin{equation} 
\label{25}
w\sim\exp\left(-A_{0}-A_{1}\right),
\end{equation}
where the parts of the total action are determined by Eqs.~(\ref{20}) and (\ref{24}). It is not difficult to check that the 
trajectory result (\ref{25}) corresponds to the top of the curve 1-2 in Fig.~\ref{fig2}(b). The trajectory is marked as the
dashed curve in Fig.~\ref{fig2}(b).

The trajectories, which are reflected from the caustics in Fig.~\ref{fig1}, can be obtained from formulas analogous to
Eqs.~(\ref{22}). The trajectory, reflected from the caustic at the physical point $\{x_{c},y_{c}\}$, has the form
\begin{equation}
\label{26}
x=x_{c}+\frac{(y-y_{c})^{2}}{4(x_{c}-1)},
\end{equation}
where $x$ is real and $(y-y_{c})$ is imaginary.
\section{DISCUSSIONS}
As shown in Fig.~\ref{fig2}, the trans-barrier state is a static packet with the top following the classical trajectory 
(\ref{11}). The associated current is directed towards the barrier at $y<0$ and is opposite at $0<y$. The $y$ component 
of the current is directed in the same way as in the initial state in the wire. The variable amplitude of the $\delta$ 
potential in Eq.~(\ref{3}) is equivalent to the effective potential $-1-\alpha^{2}(y)$ in the $\delta$ well, where the 
state of a restricted electron is a superposition of ones with two opposite velocities. Accordingly, the trans-barrier 
state is also the analogous superposition. 

The trans-barrier packet is generated at the points $\{x_{c},\pm y_{c}\}$. At a larger $|y|$ it smears out according to 
the Hamilton-Jacobi equation but keeping constant amplitude along the classical trajectory. If to account for quantum 
effects, beyond the  Hamilton-Jacobi equation, its amplitude reduces at a larger $|y|$. 

At $y_{c}<|y|$ the trans-barrier packet and the initial branch 1-1 in Fig.~\ref{fig2}(a) are not connected. In contrast, 
at $|y|<y_{c}$ the initial branch softly (without a violation of the semiclassical conditions) undergoes into the 
trans-barrier packet, the curve 1-2 in Fig.~\ref{fig2}(b). 

Suppose the system to be artificially kept to prevent a generation of the trans-barrier packet. After being released the 
system generates the packet within the short (nonsemiclassical) time of the order of $\hbar/u_{0}$. This happens because 
the packet is a superposition of all eigenstates in the $x$ channel (at a finite $x$ the variables are separated) 
including overbarrier ones. Due to that a matrix element of the initial and trans-barrier states is not exponentially small. 
One can say that the system exhibits an instability with respect to generation of the trans-barrier state. We do 
not study details of a dynamical process of filling out of the space at $x>x_{0}$ resulted from the instability.

The above scenario contrasts to the conventional resonant tunneling (Wigner tunneling), when two levels at two sides of a 
barrier are about to coincide \cite{LANDAU}. In this case the instability time is exponentially long since the matrix element 
of the states, belonging to the left and right wells, is exponentially small.

The phenomenon of generation of a large amplitude trans-barrier state is possible if the underbarrier state has a stiff 
connection with a state in the initial well. This occurs in a thin wire or a film with a transverse quantization of levels. 
In the opposite limit of an almost classical well, from where tunneling occurs, a scenario becomes conventional. It relates 
with the extreme path orthogonal to a border of the classically allowed region, where all directions of a velocity are 
possible. 

The enhanced penetration through a classical barrier can be observed in experiments for tunneling from a thin wire or a 
film with a transverse quantization of levels. An inhomogeneity of the samples should satisfy the certain not very rigorous 
conditions of the type described in Sec.~\ref{sol}.
\section{CONCLUSIONS}
 
Quantum tunneling from a thin wire or a thin film through a static potential barrier in a zero magnetic field is studied. 
The wire or the film should satisfy a condition of transverse quantization of levels and be inhomogeneous. Depending on a 
form of the inhomogeneity the tunneling scenario can dramatically differ from the conventional scheme of an exponential 
decay of a probability density inside the barrier. This happens due to interference of various undrebarrier paths, which 
are collected at some points giving rise to a trans-barrier state of a large amplitude. As a result, the tunneling 
probability through an almost classical barrier can be not exponentially small. This is a phenomenon of Euclidean resonance 
studied earlier for tunneling across nonstationary barriers.

\end{document}